# Building and exploring libraries of atomic defects in graphene: scanning transmission electron and scanning tunneling microscopy study

Maxim Ziatdinov[1,2,3], Ondrej Dyck[2,3], Bobby G. Sumpter[1,2,3], Stephen Jesse[2,3], Rama K. Vasudevan[2,3], Sergei V. Kalinin[2,3]

Affiliations:

[1]*Computational Sciences & Engineering Division, Oak Ridge National Laboratory, Oak Ridge TN 37831*

[2]*Center for Nanophase Materials Sciences, Oak Ridge National Laboratory, Oak Ridge TN 37831*

[3]*Institute for Functional Imaging of Materials, Oak Ridge National Laboratory, Oak Ridge TN 37831*

Population and distribution of defects is one of the primary parameters controlling materials functionality, are often non-ergodic and strongly dependent on synthesis history, and are rarely amenable to direct theoretical prediction. Here, dynamic electron beam-induced transformations in Si deposited on a graphene monolayer are used to create libraries of the possible Si and carbon vacancy defects. Automated image analysis and recognition based on deep learning networks is developed to identify and enumerate the defects, creating a library of (meta) stable defect configurations. The electronic properties of the sample surface are further explored by atomically resolved scanning tunneling microscopy (STM). Density functional theory is used to estimate the STM signatures of the classified defects from the created library, allowing for the identification of several defect types across the imaging platforms. This approach allows automatic creation of defect libraries in solids, exploring the metastable configurations always present in real materials, and correlative studies with other atomically-resolved techniques, providing comprehensive insight into defect functionalities. Such libraries will be of critical importance in automated AI-assisted workflows for materials prediction and atom-by atom manipulation via electron beams and scanning probes.

**Introduction.**

Materials form the basis of modern civilization, ranging from structural materials in buildings and machines to semiconductors underpinning information technologies to functional materials in batteries, fuel cells, sensors and other applications. For many materials classes, functionality is ultimately determined by the presence and configurations of defects, ranging from simple substitutional defects in solid solutions to complex extended defects involved in crystallographic shear phases, dislocations, etc.

Of course, the use of high throughput computations to create libraries of materials is not new, as exemplified by Materials Genome Initiative.[1] In this, large data bases of first principle calculations were created and subsequently mined for materials with interesting properties or for establishing structure-property relationships. However, while powerful for the simple materials, the number of possible structures for extended defects makes direct computational search infeasible. An alternative approach for materials design is based on artificial intelligence-assisted experimental workflows, where thermodynamic data bases, DFT calculations, etc. are integrated to achieve property predictions and establish synthesis-property relationships. The corresponding data bases include Materials Data Facility,[2] Citrination,[3] Dark Reactions,[4] Materials Innovation Network.[5] However, to date these efforts utilize macroscopic experimental descriptors and mesoscopic structural descriptors only. Very recently, Rajan *et al.*[6] created a theoretical library of nanosized pore defects in 2D materials and explored their energetics.

Given advances in imaging, it is now possible to image many systems with atomic resolution via electron or scanning probe microscopy.[7-12] The key to understanding and utilizing this data is to extract the different configurations of the atomic species, as their distributions (fluctuations) encode important aspects that govern the system's response to thermodynamic perturbations. However, this requires locating and classifying the defects themselves in an automated manner, due to the numbers required for proper statistics. The task is even more important given the need for complementary data on the same defect configurations. For instance, imaging of atomic structure of defects is relatively straightforward in scanning transmission electron microscopy (STEM). However, the finer details of electronic structure at or around the defect necessitate scanning tunneling microscopy (STM) measurements. However, due to the differences in imaging mechanisms and limitations of the experimental platforms, correlative atomically resolved STEM-STM studies are extremely complex. Furthermore, the contrast in STM and STEM data is completely different, so the complementary nature of both techniques cannot always be readily utilized. What is needed, then, is a combined experimental-theoretical approach to extract defects, classify them, and then perform first-principles calculations to explore their electronic structure. The latter can immediately be used to simulate the expected STM images, and provide direct insight into the effects of the defect on the material's properties.

Here we demonstrate an experimental approach for creating defect libraries from scanning transmission electron microscopy data and determine defect statistics in the Si-graphene system via advanced image recognition tools. We further use these as an input for the density functional structure theory calculations to establish associated electronic properties and compare these with the scanning tunneling microscopy results on the same system.

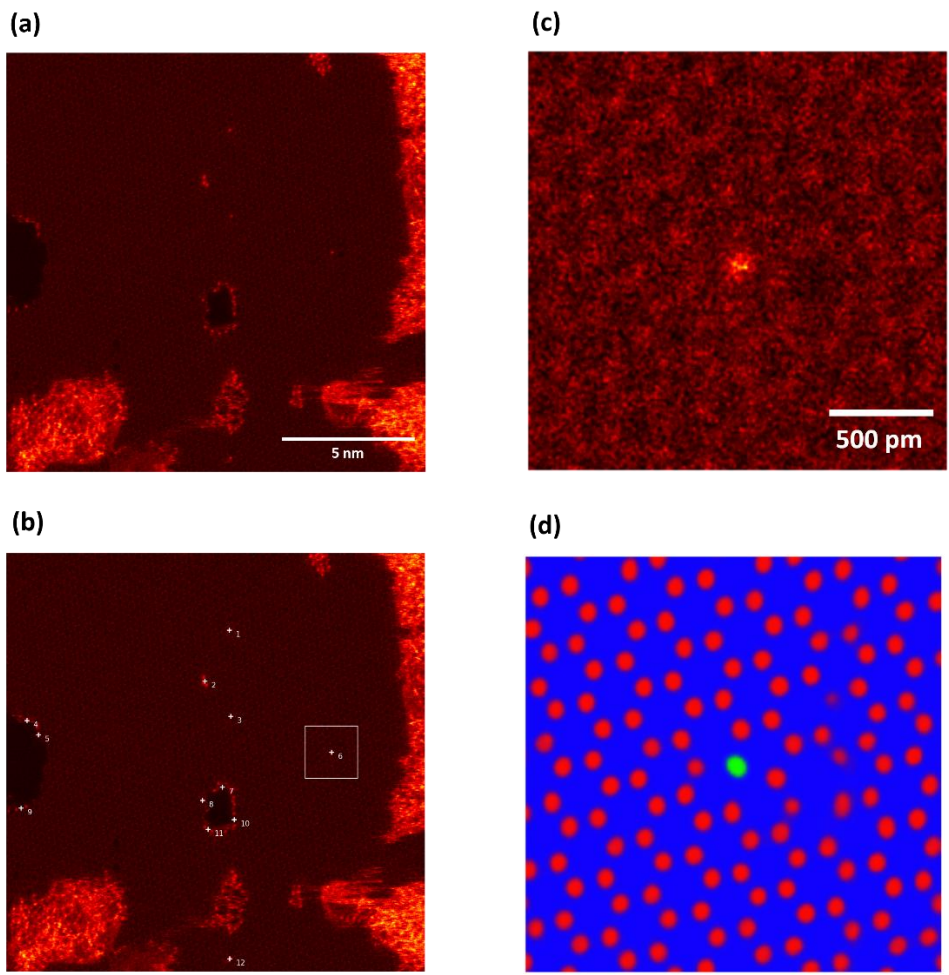

**FIGURE 1.** (a) STEM overview image. Brighter areas typically correspond to amorphous Si-C regions, from which Si atoms can be dispersed into clean graphene lattice, leading to various Si

defect structures (in terms of number of Si atoms and their bonding to each other and to nearby lattice atoms). (b) Results of applying a "defect sniffer" (deep neural network followed by density based spatial clustering) to data in (a). White markers show regions of interest as identified by a "defect sniffer". Notice that it ignores amorphous Si-C regions and returns information on the location of point Si defects only (c) 2 nm × 2 nm area cropped around one of the identified defects (white box in (b)). This cropped image serves as an input to "atom finder" model. (d) The output of an "atom finder" network for the image in (c), where red blobs correspond to C lattice atoms and a green blob is associated with Si impurity.

**Results and Discussion.**

To create libraries of defects in graphene, we used active manipulation via 100 kV electron beam to disperse silicon (Si) impurities over the graphene lattice.[13,14] This active sputtering approach allowed us to create multiple metastable configurations of Si-C complexes in graphene. We then acquired images using a 60 kV electron beam, which minimizes damage to the graphene lattice and allows a closer inspection of the formed defect structures. Fig. 1a shows an example high angle annular dark field (HAADF) STEM image of a graphene sample with amorphous Si-C regions which serve as the source of Si dopant atoms. Individual Si impurities embedded in the graphene lattice are seen.

To analyze the created defect structures and to generate a library of defects, we have developed a deep-learning-based analytical workflow that searches for and identifies various defect configurations in the available "raw" experimental data and categorizes them based on the number of Si atoms in each configuration. As a first step, we employed two deep fully convolutional neural networks (see Methods for details). The goal of the first neural network, a "defect sniffer", is to identify anomalies and irregularities (that is, defects) in the periodic structure of a crystalline material (in this case, graphene). The second neural network, an "atom finder", identifies the elemental species and positions of atoms in a localized region (typically 2 nm by 2 nm) around the detected defect. This two-stage deep learning approach resembles the logic of a human operator who first identifies points of interest (irregularities and "anomalies") in a larger scale experimental image and subsequently examines specific features (e.g. details of atomic structure surrounding the detected irregularity) in more detail. Thus, it can also potentially be applied in future for a realization of intelligent ("self-driving") microscope. It is worth noting that while this two-stage approach can be in principle replaced by just a single deep learning model, the classical generalization *vs.* accuracy tradeoff will likely increase the error of finding positions of atoms and atomic defects. We therefore believe that the current approach (with two specialized models rather than one very general model) is an optimal solution for this type of problem.

The "defect sniffer" neural network was trained using several weakly-labeled experimental images, utilizing the fact that each defect is associated with violation of ideal periodicity of the lattice. The details of the procedure for constructing a training set for this type of problem can be found elsewhere.[15] Importantly, it was trained to find point defects only (vacancies and Si dopants), while avoiding areas with extended defects (such as larger holes and amorphous Si-C regions). Density-based spatial clustering analysis[16] was applied to the output of a neural network to group multiple detected point defects located close to each other (maximum radius $\varepsilon$ set to 0.5 nm). The regions of interest (in terms of presence of Si point defects) identified by this approach for data from Fig. 1a are shown in Fig. 1b.

The "atom finder" neural network was trained using simulated STEM images. Recently, several studies[17-19] have demonstrated that deep neural networks trained using simulated data can successfully generalize to real experimental atom-resolved and molecule-resolved scanning probe and transmission electron microscopy data. The theoretical data was produced by either using a MultiSlice algorithm[20] for STEM image simulations or by representing atoms as Gaussian blobs. Interestingly, both approaches for model training gave similar results in term of defect finding and identification for the current system. Each theoretical image was augmented to account for model uncertainties such as local distortions (± 15% displacements in atomic positions) and instrumental factors such as global image/scan distortion and different levels of noise.

The output of the second neural network was mapped onto the graph structure representing the classical chemical bonding picture.[19] Specifically, we first identified the centers of the mass for each "blob" (atom) in the pixel-wise maps from the output(s) of a neural network and then use the information about the type and position of each atom to construct the graph nodes. The nodes were then connected by the graph edges which represent bonds. To construct the bonds, we use chemistry-based hard constraints on the maximum possible coordination number of each type of atom, as well as maximum/minimum allowed length of atomic bond between corresponding pairs of atoms. We note that this analysis assumes that a sample has a quasi-2D configuration. If there is no information (e.g. from image "tags") available on the scan size, it is possible switch to calibration based on the peak in the distribution of nearest neighbor distances (which are predominantly C-C bonds).

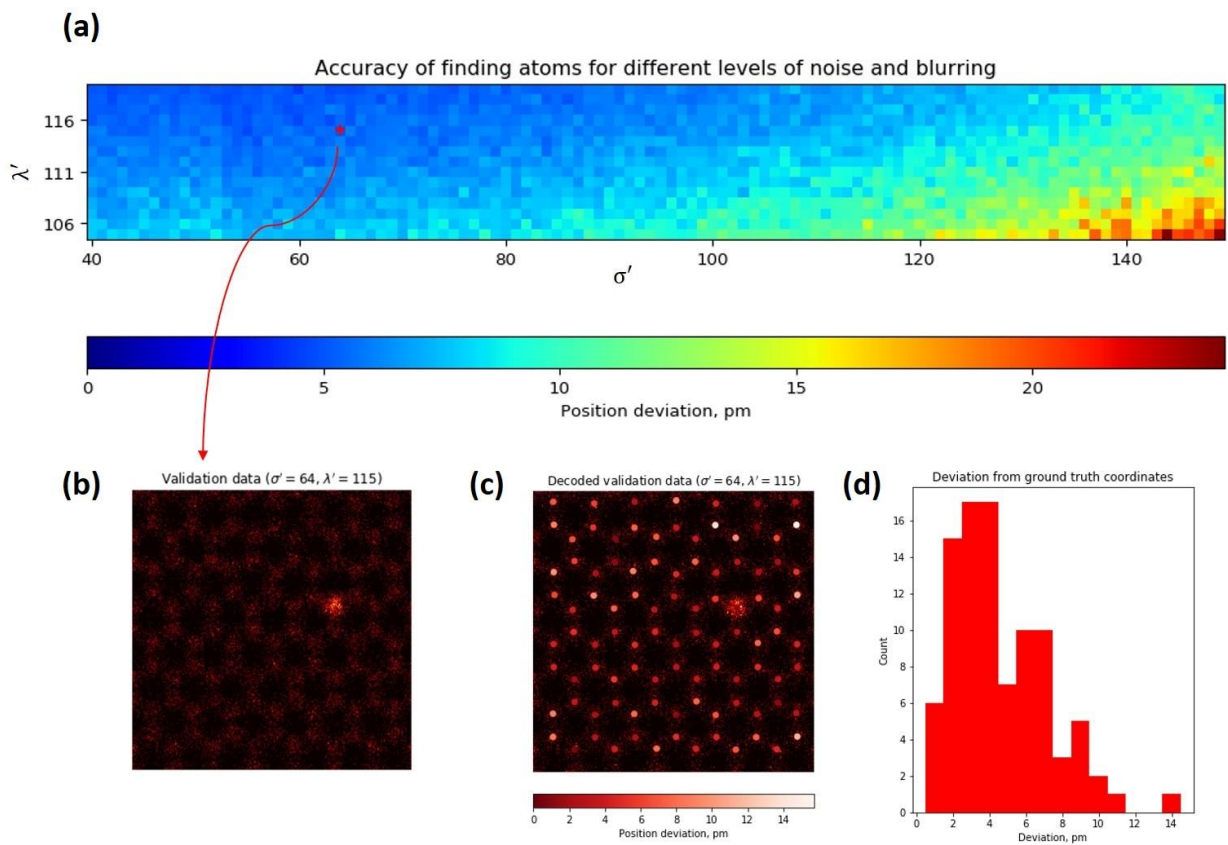

**FIGURE 2**. **Evaluation of accuracy in finding the atomic positions via a deep learning model for different levels of noise.** (a) The deviation of the predicted positions (average value for all atoms in the image) from the "ground truth" positions of atoms for different combinations of noise ($\lambda'$) and blurring ($\sigma'$). The $\lambda'$ and $\sigma'$ are the scaled parameters of Poisson and Gaussian distributions, respectively. See supplemental material for the details and code of how the noisy data was created and analyzed. (b) Simulated image of graphene lattice associated with red cross in image in (a). (c) Same image overlaid with atomic coordinates. Different colors of circular markers show different degree of atomic positions displacement from "ground truth" coordinates. (d) The histogram showing atomic displacements for every atomic position displayed in (c). Most of the deviations are below 10 pm, which is within instrumental uncertainty for atomic position extraction.

Here it was crucial to identify accurately the environment of each Si impurity, which included the information on the lengths and angles of bonds with the neighboring carbon atoms in the host graphene lattice. To the best of our knowledge, the current methods for atom finding in atomically resolved data does not allow a reliable extraction of atomic positions for images where atoms do not appear as local maxima (such as in Fig. 1c). As a result, there is no ground truth for the experimental data against which our deep learning based "atom finding" approach can be tested. Hence, we used images of graphene lattice simulated with MultiSlice algorithm that were not a part of the model training/testing and therefore can serve as a validation dataset. The number of atoms in the validation images was comparable to that in the experimental images used in the current study (in the second, "atom finding", stage). Each image was corrupted by unique combination of blurring and noise (no two images were the same). The average (per each image) deviation of the predicted positions from the "ground truth" positions of atoms for different combination of noise and blurring is shown in Fig. 2a. We generally found that the trained model can determine robustly atomic positions up until the level of noise becomes so high that the lattice contours become barely distinguishable for a human eye (this corresponds to regions with $\sigma' > 130\ a.u.$ and $\lambda' < 110\ a.u.$ in Fig. 2a). In Fig. 2b-d we showed an example of the model output for data where the image resolution, number of atoms in image and the level of noise are comparable to those in current experiments. The deviation of atomic center positions from "true" positions for such data is mostly below 10 pm, which is within the instrumental error for detection of atomic position in single image STEM data with low signal-to-noise-ratio.[21,22] We believe that because getting the atomic positions accurately is crucial for this particular type of pixel-wise classification problem, the approach for testing a network's accuracy described above is preferable to the existing ones from the domain of classical machine/deep learning.

The total number of analyzed images was approximately 500, with each image containing either single or multiple Si impurities. The image size varied between 2 nm and 16 nm and the image resolution was in the range between 256 px × 256 px and 2048 px × 2048 px. It is worth noting that many of the same types of Si defects can exist without 100 kV e-beam "sputtering", but with a much smaller density, which is not sufficient for creating a library of defects. The representative examples for each category were then selected with the assistance of a human expert. Structures which could not be confirmed independently (even on a simple qualitative level) by a human expert eye, either due to very high levels of noise in the image or due to ambiguity associated with absence of information in the z-direction for certain structures, were not considered/counted. Finally, the input (structural) data for the first-principles-based studies of the detected defect were obtained by cropping a square "cell" with ~30 atoms around the defect of interest (for the chosen categories) and embedding them into a larger standard graphene supercell.

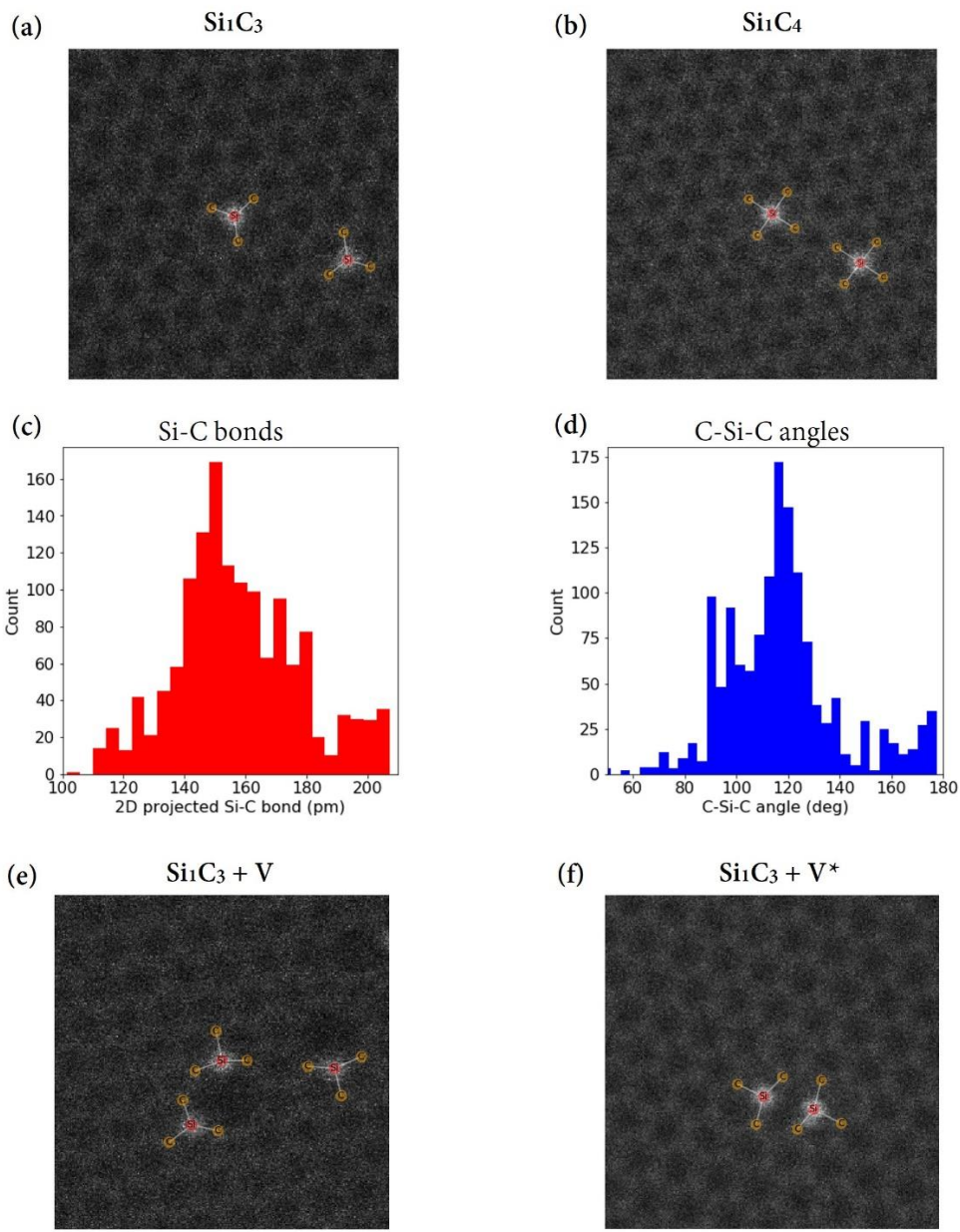

**FIGURE 3. Constructing libraries of defects from STEM data on graphene with Si impurities via deep-learning-based analysis of raw experimental data.** (a-b) Defects containing a single 3-fold coordinated Si atom (count: 465) (a) and a single 4-fold coordinated Si atom (count: 16) (b). (c, d) Histograms of Si-C bond lengths (c) and Si-C bond angles (d) for 3-fold Si defect. (e, f) Examples of distorted 3-fold Si defect located next to a multi-vacancy (e) and at topological defect (f). The number of distorted ("standard deviation" $s = (\sum dx^2/n)^{1/2}$ in C-Si-C angles above 15) and undistorted 3-fold Si defect was 231 and 234, respectively.

Figure 3-5 illustrate a constructed library of Si-containing defects in graphene. We define a Si defect as a Si atom and its first coordination "sphere" or Si atoms connected either directly or through a single C atom to each other and their first coordination "sphere". Using this definition of defect, we began categorizing Si defects based on the number of Si atoms that each defect contained. The defects containing a single Si and two Si are shown in Fig. 3 and Fig. 4, respectively. The defects containing three or more Si impurities are shown in Fig. 5. Our approach also allowed a further split of these categories into subcategories based on coordination number, variation in bond lengths in a single defect, proximity to larger defects (e.g. nanoholes), etc. It is worth noting that since our analysis was performed after the STEM experiments were completed, it does not account for the possibility that certain defects appear more frequently than others simply due to human operator (confirmation) bias and specific interest in certain structures. Hence, the frequency of different defect occurrences in the analyzed images may not necessarily reflect the relative frequency of their occurrence in the physical system. However, automatization of experimental acquisition will preclude these problems in the future.

The first category is a defect that has only one Si atom (Fig. 3). It contains two subcategories, based on number of bond connections to nearby lattice atoms, corresponding to classical 3- and 4 coordinated Si atoms (Fig. 3a and 3b, respectively).[23] Our model was able to identify 465 three-fold Si atoms and 16 four-fold coordinated Si atoms. This approach also enables an automated calculation of the bond lengths and bond angles in each detected defect. The histograms of bond lengths and angles for all the detected 3-fold Si defect are shown in Fig. 3c and 3d, respectively. We next defined a "distorted" 3-fold defect as a defect where a "standard deviation", $s = (\sum dx^2/n)^{1/2}$, of three C-Si-C bond angles is larger than 15. This allowed the identification of additional subcategories of 3-fold defect corresponding to Si impurities located adjacent to hole-multi-vacancies (Fig. 3e) or to a topological defect (Fig. 3f), which are characterized by a significantly stronger distortion of C-Si-C bond angles (Note that both missing atoms in honeycomb lattice and topological reconstructions with enlarged size of carbon rings are seen as "vacancies" in our analysis).

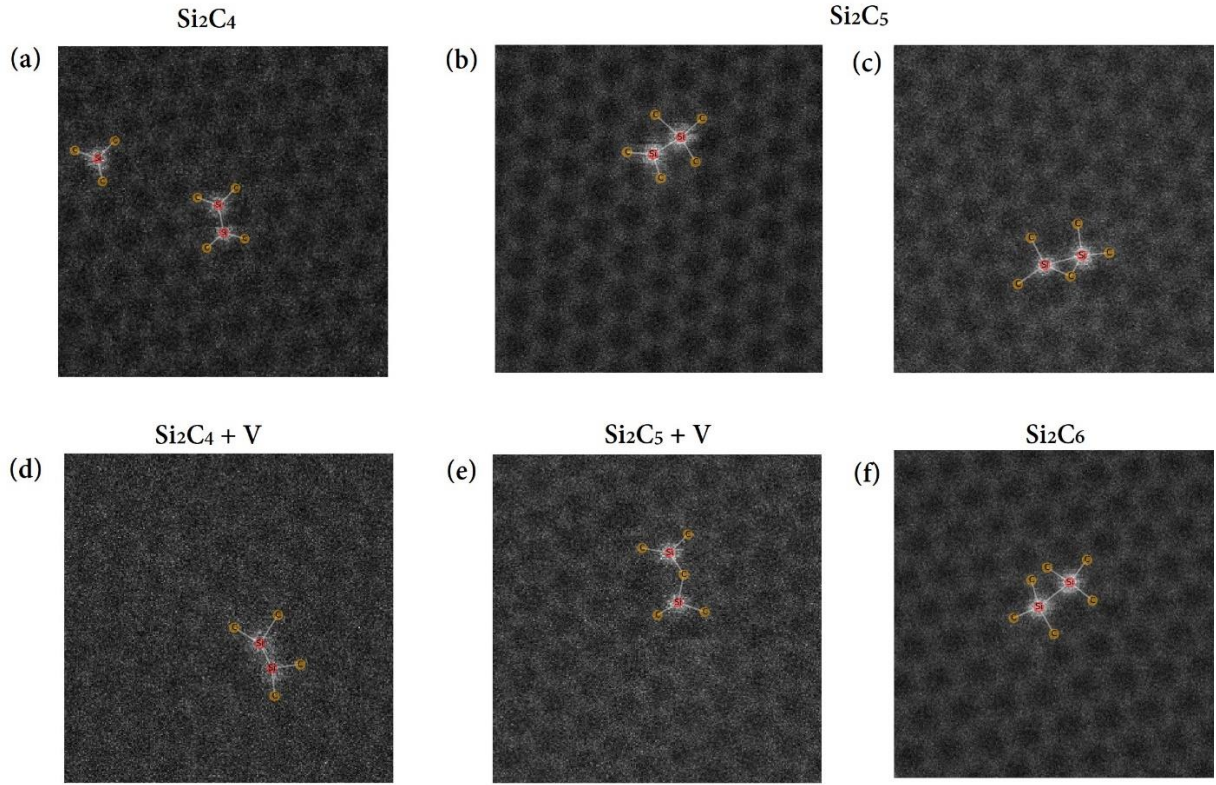

**FIGURE 4. Atomic defects containing two Si atoms.** Si dimer structures with each Si having 2 C in its first coordination sphere (count: 6) (a, d); one Si having 3 C atoms and another one having 2 C atoms (count: 4) (b); each Si having 2 C atoms plus one "shared" C atom (count: 3) (c) ; each Si connected to 3 C atoms (count: 5) (f); non-dimer structure, where two Si atoms are connected via C atom (count: 3) (e).

We proceed to discussion of defect containing more than a single Si atom. Figure 4 shows the identified structures of 2-Si defect. This includes multiple different dimer structures, in which each Si atom in the dimer has 3-fold or 4-fold coordination (Fig. 4a-d, f), as well as a 2-Si defect where Si atoms are connected to each other via a C atom (Fig. 4e). As in the case with a single atom, the 2-Si defects located next to the hole/multi-vacancy, are characterized by a stronger distortion in C-Si-C bonds compared to when they are observed on a clean lattice (see Fig. 4a and 4d). A third category includes Si defects with three or more Si atoms (Fig. 5). This category contains trimer (Fig. 5a) and tetramer (Fig. 5b) arrangements of Si impurities as well as more complex Si-C clusters containing five and more Si that are usually observed at the edges of graphene vacancies and holes.

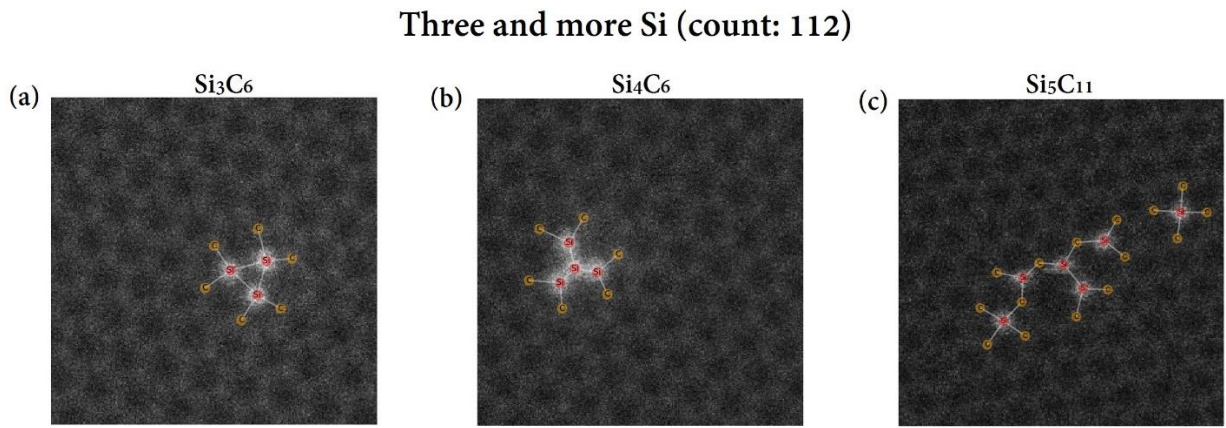

**FIGURE 5. Atomic defects containing three or more Si atoms.** Constructing libraries of defects from STEM data on graphene with 3 Si (count: 50) (a), 4 Si (count: 6) (b) and 5 or more Si (count: 56) (c).

This comprises the library of observed Si atomic defects in monolayer graphene. We made the images labeled according to the classification scheme described above available through Citrination platform.[24] To demonstrate the applicability of this library for material exploration and additional insight into the electronic structure of defects in graphene with Si, we perform density functional theory (DFT) based calculations of electronic structure as well as experimental atomic-scale scanning tunneling microscopy (STM) measurements on the same sample. The STM images obtained at low bias voltages reflect spatial distribution of electronic states around the Fermi level and can be used to probe local perturbation in a material's electronic structure caused by defects. An example of an experimental STM image obtained over a relatively large field of view on the sample from STEM measurements is shown in Figure 6. The STM data typically showed the presence of atomically-resolved regions with a lateral size of about 3 – 6 nm separated by regions for which no atomic resolution could be achieved. We assigned the latter to amorphous C/Si regions, which were commonly seen in the STEM experiment on this and other samples. An example of a point defect causing perturbation in the electronic structure of a nearby graphene region is shown in the magnified image in Fig. 6b.

It is important to note two important limitations expected in the interpretation of STM images from a STEM sample. First, because the sample was transferred through ambient atmosphere, contamination of the sample surface with oxygen functional groups is possible. In addition, in the STM setup used for this study it was not possible explore the exact same region that was examined in the STEM experiment.

Despite these limitations, we were able to capture a number of atomic-scale STM images with characteristic point defects and compare them to the structures derived from the analysis of the STEM data. Note that a proper interpretation of STM data requires comparison with theoretical calculations of local density of states at the Fermi level for the structures of interest. For this purpose, we performed DFT calculations for the first two categories of defects. We limited ourselves only to single and dimer Si defects occurring on a clean graphene surface. Specifically, we first performed full 3D relaxation of the 2D projection of atomic coordinates found in the analysis of the STEM data. Then, the relaxed coordinates were used to calculate the electronic structure for each type of defect. The STM data for point defects was then compared with DFT-calculated STM images.

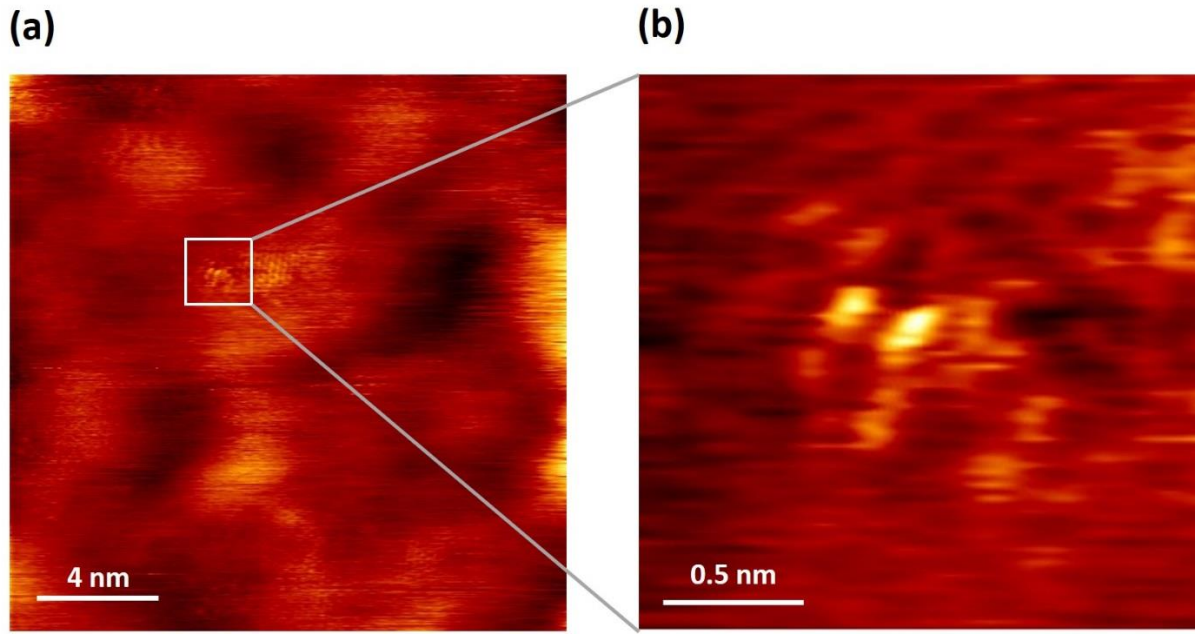

**FIGURE 6.** (a) Large-scale STM image of a relatively large area of the graphene sample from the STEM experiment. (b) Magnified view of an atomic defect highlighted by the white rectangle in (a). Tunneling conditions $V_{bias}$ = 0.1 V, $I_{setpoint}$ = 55 pA.

We compared the experimental STM data with DFT calculations of integrated local density of states associated with two bands above and below the Fermi level (if such bands existed within -0.5 eV to +0.5 eV energy range). We note that of the structures analyzed only the 3-fold Si impurity showed an out-of-the-plane displacement in the final relaxed geometry, whereas the rest of the structures remained flat. Comparison of point-like defects obtained in the STM at low-bias voltage (0.1 V to 0.3 V) to simulations of 3-fold and 4-fold Si defect (Fig. 7) suggests that we were able to capture the former in the STM experiment. We next compared the dimer-like structures observed in the STM experiment (Fig. 8) to the DFT simulations of dimer structures observed in Fig. 4. We found that the closest match is the $Si_2C_6$ (Fig. 4f) structure that is characterized by two well-defined bright spots associated with two Si impurities in the relevant energy range. Indeed, the rest of the tested dimer structures either did not produce a well-defined structure with two bright spots or the distance between the spots was smaller than in the case of STM experimental data (see Supplementary Information). Finally, in Fig. 9 we show two additional structures which are possible candidate matches with those displayed in Fig. 5, containing three (Fig. 9a) and more than three (Fig. 9b) Si atoms. The interpretation of the exact nature of such complex defects will be the subject of future work.

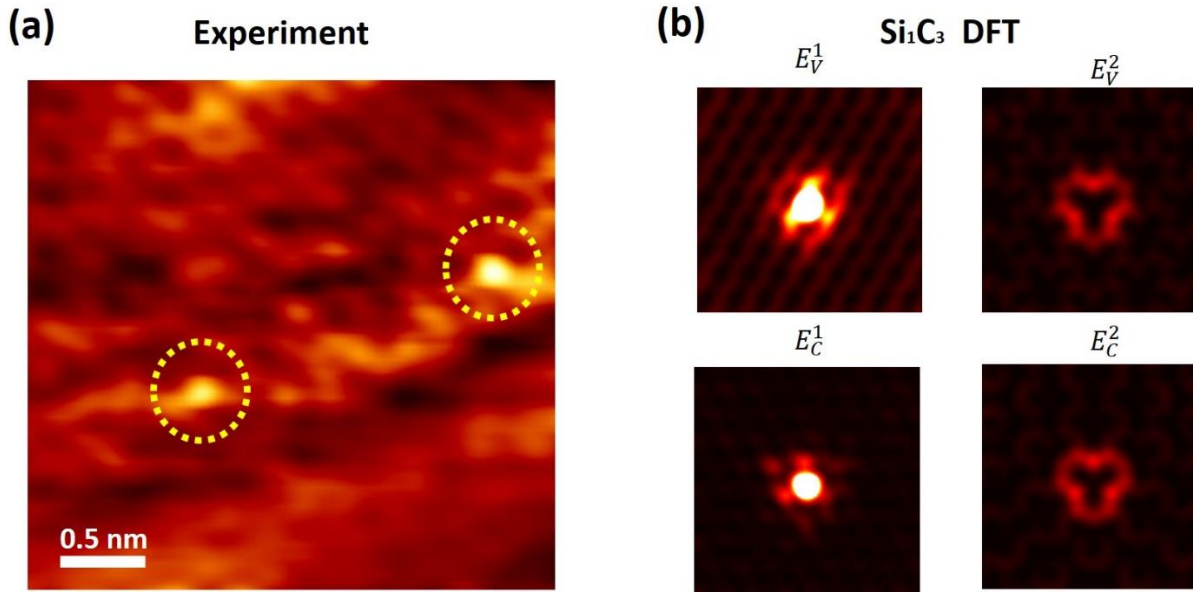

**FIGURE 7. Comparison of STM data on point impurities with DFT calculations.** (a) Experimental STM data obtained at tunneling conditions $V_{bias} = 0.1$ V, $I_{setpoint} = 55$ pA. (b) DFT-simulated STM images of $Si_1C_3$ defect for two bands below ($E_V^1$ and $E_V^2$) and above ($E_C^1$ and $E_C^1$) the Fermi level.

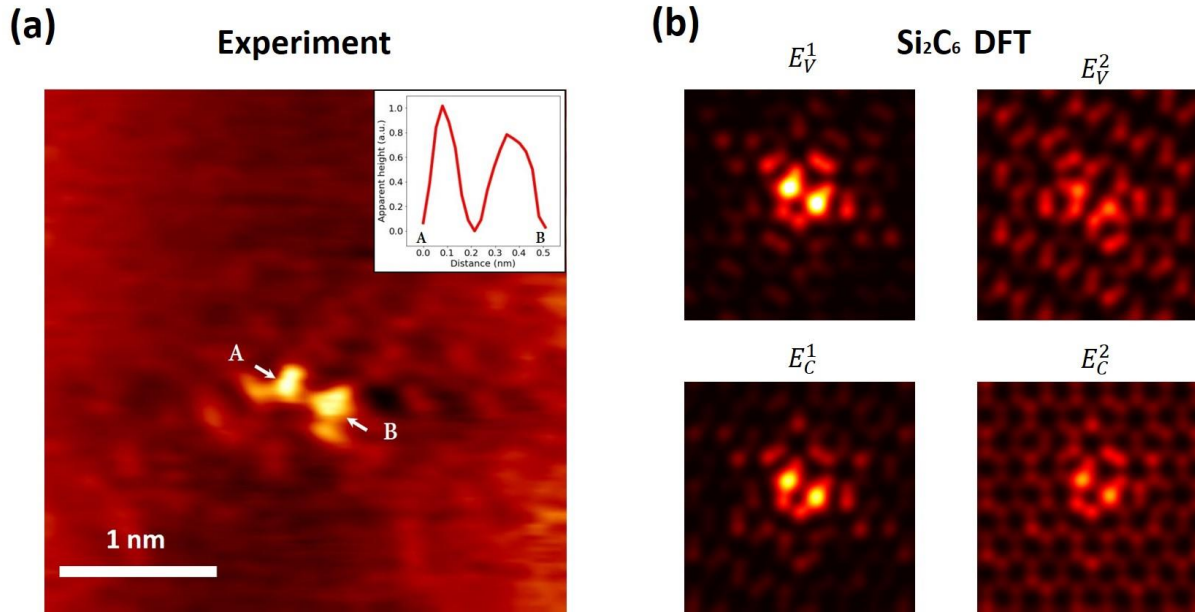

**FIGURE 8. Comparison of STM data on point impurities with DFT calculations.** (a) Experimental STM data obtained at tunneling conditions $V_{bias} = 0.3$ V, $I_{setpoint} = 55$ pA. Inset shows a line profile along between the A and B points. (b) DFT-simulated STM images of $Si_2C_6$ defect for two bands below ($E_V^1$ and $E_V^2$) and above ($E_C^1$ and $E_C^1$) the Fermi level.

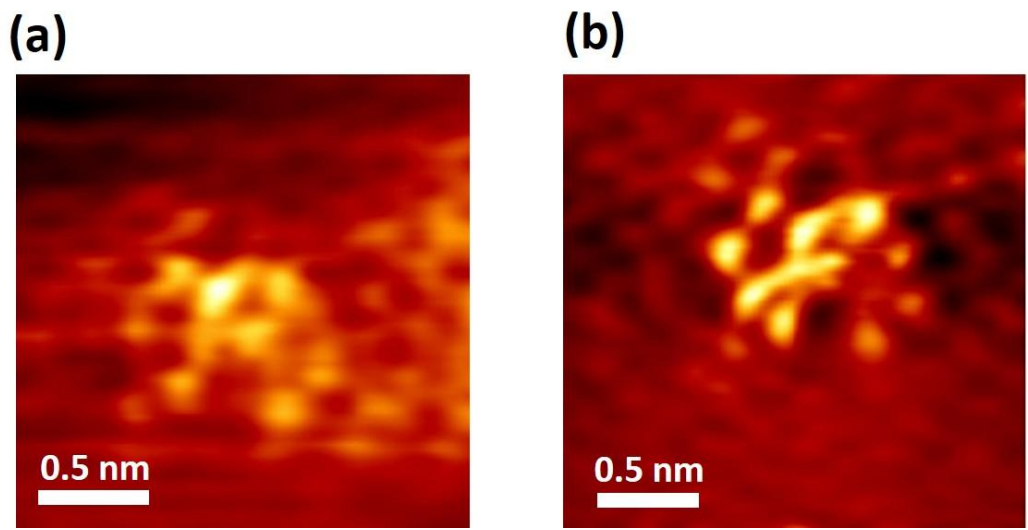

**FIGURE 9. More complex defect structures observed in the STM experiment on STEM sample of graphene.** (a, b) Examples of defect structures seen in STM that can be potentially

associated with more complex structural defects seen in STEM including defects-vacancy complexes with 3 or more Si atoms.

**Conclusions and Outlook.**

To summarize, we combined active electron beam manipulation of impurities in graphene and deep machine learning to construct libraries of Si defect structures. By performing scanning tunneling microscopy experiments on the same sample, supplemented by first principles calculations, we were able to link together experimental results that characterize a material's structure (STEM) and functionalities (STM). Our work shows a pathway toward the creation of comprehensive libraries of atomic defects and their functionalities based on experimental observations from multiple atomically-resolved probes generating synergistic information and performing correlative analysis of structure-property relationships on the level of single atomic defects.

We propose that an important part of both theoretical and experiment driven materials prediction will be libraries of atomic defects. For the theoretical effort, these libraries can significantly confine the region of the chemical space to explore by focusing the effort on the experimentally observed defect classes, as opposed to all those theoretically possible. For experiment, the defect populations and correlations can be used as direct input to machine learning schemes. Finally, this information can shed light on defect equilibria, solid state reaction pathways, and other fundamental parameters of materials.

**Acknowledgments:**

This research was sponsored by the Division of Materials Sciences and Engineering, Office of Science, Basic Energy Sciences, US Department of Energy (RVK and SVK). Research was conducted at the Center for Nanophase Materials Sciences, which is a DOE Office of Science User Facility. STEM experiments (OD, SJ) and data analysis (MZ) were supported by the Laboratory Directed Research and Development Program of Oak Ridge National Laboratory, managed by UT-Battelle, LLC, for the U.S. Department of Energy.

## Methods:

### Sample Preparation

Graphene was grown via atmospheric pressure chemical vapor deposition (APCVD) on a Cu foil.[25] The Cu foil was then spin coated with poly(methyl-methacrylate) (PMMA) as a mechanical stabilizer. The Cu foil was dissolved in a bath of ammonium persulfate-deionized (DI) water (0.05 g/ml). The graphene/PMMA film was transferred to hydrogen chloride (HCl) diluted in DI water bath to remove the ammonium persulfate, followed by a DI water rinse. The graphene/PMMA film was placed on a TEM grid and annealed on a hot plate at 150 °C for ~20 minutes to adhere the grid to the graphene. The PMMA was subsequently dissolved away in an acetone bath, followed by an isopropyl alcohol rinse. Finally the grid was annealed in an Ar-$O_2$ (450 sccm/45 sccm) environment at 500 °C for 1.5 hours to prevent e-beam induced hydrocarbon deposition in the STEM.[26,27]

### STEM experiment

STEM imaging was performed using a Nion UltraSTEM U100 STEM operated at 60 kV. The images were acquired in high angle annular dark field (HAADF) imaging mode and were introduced to the DL network without any post processing.

### STM experiment

The STM experiments were performed at room temperature using an Omicron VT-STM equipped with a Nanonis controller in an operating pressure of $2 \times 10^{-10}$ Torr. The STM images were obtained with mechanically cut Pt/Ir alloy tips.

### DFT calculations

The DFT calculations were carried out using the Vienna *Ab Initio* Simulation Package (VASP, 5.4.1)[28,29] using the projector-augmented wave (PAW) method[30,31]. The electron-ion interactions were described using standard PAW potentials, with valence electron configurations. A kinetic energy cutoff on the plane waves was set to 400 eV and the "accurate" precision setting was adopted. A Γ-centered k-point mesh of 9x9x1 was used for the Brillouin-zone integrations. The convergence criteria for the electronic self-consistent loop was set to $10^{-4}$ eV.

**Data analysis**

The deep neural networks were implemented using Keras deep learning library (https://keras.io/). The neural networks had an encoder-decoder structure (SegNet-like architecture[32]) that allowed a pixel-level classification of input images. For the "atom finder" network, the encoder part consisted of 6 convolutional layers, all activated by rectified linear unit function. The convolutional filters (kernels) in all the layers were of the size 3×3 and stride 1. The 1st convolutional layer had 64 filters, the 2nd and 3rd layers had 128 filters each, and the 4th, 5th and 6th layers had 256 filters each. The max-pooling layers were placed after the 1st, 3rd and 6th layers. The decoder part contained the same blocks of convolutional layers in reversed order with a bilinear interpolated upsampling instead of the max-pooling units. The Adam optimizer[33] was used with categorical cross-entropy as the loss function. For the "defect sniffer" network, the encoder part consisted of 3 convolutional layers with 64, 128 and 256 filters of the size 3×3 and stride 1, activated by rectified linear unit function. The max-pooling layers were placed in between the layers. The decoder part contained the same blocks of convolutional layers in reversed order with the nearest-neighbor upsampling between them. The focal loss[34] was used for improving identification of the defect structures. To optimize this loss, the Adam optimizer was used. The accuracy score on a test set for "atom finder" and "defect sniffer" were ≈96% and ≈99%, respectively. The graph structures were constructed using NetworkX library (https://networkx.github.io/).

# References

(1) Green, M.; Choi, C.; Hattrick-Simpers, J.; Joshi, A.; Takeuchi, I.; Barron, S.; Campo, E.; Chiang, T.; Empedocles, S.; Gregoire, J. Fulfilling the promise of the materials genome initiative with high-throughput experimental methodologies. *Applied Physics Reviews* **2017**, *4*, 011105.

(2) Blaiszik, B.; Chard, K.; Pruyne, J.; Ananthakrishnan, R.; Tuecke, S.; Foster, I. The Materials Data Facility: Data services to advance materials science research. *JOM* **2016**, *68*, 2045-2052.

(3) O'Mara, J.; Meredig, B.; Michel, K. Materials data infrastructure: a case study of the citrination platform to examine data import, storage, and access. *JOM* **2016**, *68*, 2031-2034.

(4) Raccuglia, P.; Elbert, K. C.; Adler, P. D. F.; Falk, C.; Wenny, M. B.; Mollo, A.; Zeller, M.; Friedler, S. A.; Schrier, J.; Norquist, A. J. Machine-learning-assisted materials discovery using failed experiments. *Nature* **2016**, *533*, 73.

(5) Kalidindi, S. R.; Brough, D. B.; Li, S.; Cecen, A.; Blekh, A. L.; Congo, F. Y. P.; Campbell, C. Role of materials data science and informatics in accelerated materials innovation. *MRS Bulletin* **2016**, *41*, 596-602.

(6) Govind Rajan, A.; Silmore, K. S.; Swett, J.; Robertson, A. W.; Warner, J. H.; Blankschtein, D.; Strano, M. S. Addressing the isomer cataloguing problem for nanopores in two-dimensional materials. *Nature Materials* **2019**, *18*, 129-135.

(7) Nellist, P. D.; Chisholm, M. F.; Dellby, N.; Krivanek, O. L.; Murfitt, M. F.; Szilagyi, Z. S.; Lupini, A. R.; Borisevich, A.; Sides, W. H.; Pennycook, S. J. Direct Sub-Angstrom Imaging of a Crystal Lattice. *Science* **2004**, *305*, 1741.

(8) Pennycook, S. J.; Nellist, P. D.: *Scanning Transmission Electron Microscopy*; Springer-Verlag New York, 2011.

(9) Krivanek, O. L.; Chisholm, M. F.; Nicolosi, V.; Pennycook, T. J.; Corbin, G. J.; Dellby, N.; Murfitt, M. F.; Own, C. S.; Szilagyi, Z. S.; Oxley, M. P.; Pantelides, S. T.; Pennycook, S. J. Atom-by-atom structural and chemical analysis by annular dark-field electron microscopy. *Nature* **2010**, *464*, 571.

(10) Wiesendanger, R.: *Scanning probe microscopy and spectroscopy: methods and applications*; Cambridge university press, 1994.

(11) Wiesendanger, R.: *Scanning probe microscopy: analytical Methods*; Springer Science & Business Media, 1998.

(12) Stroscio, J. A.; Kaiser, W. J.: *Scanning tunneling microscopy*; Academic Press, 1993; Vol. 27.

(13) Dyck, O.; Kim, S.; Jimenez-Izal, E.; Alexandrova, A. N.; Kalinin, S. V.; Jesse, S. Assembling Di- and Multiatomic Si Clusters in Graphene via Electron Beam Manipulation. *ArXiv e-prints* **2017**.

(14) Dyck, O.; Kim, S.; Kalinin, S. V.; Jesse, S. Placing single atoms in graphene with a scanning transmission electron microscope. *Appl. Phys. Lett.* **2017**, *111*, 113104.

(15) Maksov, A.; Dyck, O.; Wang, K.; Xiao, K.; Geohegan, D. B.; Sumpter, B. G.; Vasudevan, R. K.; Jesse, S.; Kalinin, S. V.; Ziatdinov, M. Deep learning analysis of defect and phase evolution during electron beam-induced transformations in WS2. *npj Computational Materials* **2019**, *5*, 12.

(16) Ester, M.; Kriegel, H.-P.; Sander, J.; Xu, X. A density-based algorithm for discovering clusters in large spatial databases with noise. *Kdd* **1996**, *96*, 226-231.

(17) Ziatdinov, M.; Maksov, A.; Kalinin, S. V. Learning surface molecular structures via machine vision. *npj Computational Materials* **2017**, *3*, 31.

(18) Madsen, J.; Liu, P.; Kling, J.; Wagner, J. B.; Hansen, T. W.; Winther, O.; Schiøtz, J. A Deep Learning Approach to Identify Local Structures in Atomic-Resolution Transmission Electron Microscopy Images. *Advanced Theory and Simulations* **2018**, *0*, 1800037.

(19) Ziatdinov, M.; Dyck, O.; Maksov, A.; Li, X.; Sang, X.; Xiao, K.; Unocic, R. R.; Vasudevan, R.; Jesse, S.; Kalinin, S. V. Deep Learning of Atomically Resolved Scanning Transmission Electron Microscopy Images: Chemical Identification and Tracking Local Transformations. *ACS Nano* **2017**, *11*, 12742-12752.

(20) Dr. Probe - High-resolution (S)TEM image simulation software. http://www.er-c.org/barthel/drprobe/.

(21) Savitzky, B. H.; El Baggari, I.; Clement, C. B.; Waite, E.; Goodge, B. H.; Baek, D. J.; Sheckelton, J. P.; Pasco, C.; Nair, H.; Schreiber, N. J.; Hoffman, J.; Admasu, A. S.; Kim, J.; Cheong, S.-W.; Bhattacharya, A.; Schlom, D. G.; McQueen, T. M.; Hovden, R.; Kourkoutis, L. F. Image registration of low signal-to-noise cryo-STEM data. *Ultramicroscopy* **2018**, *191*, 56-65.

(22) Yankovich, A. B.; Berkels, B.; Dahmen, W.; Binev, P.; Sanchez, S. I.; Bradley, S. A.; Li, A.; Szlufarska, I.; Voyles, P. M. Picometre-precision analysis of scanning transmission electron microscopy images of platinum nanocatalysts. *Nature Communications* **2014**, *5*, 4155.

(23) Ramasse, Q. M.; Seabourne, C. R.; Kepaptsoglou, D.-M.; Zan, R.; Bangert, U.; Scott, A. J. Probing the Bonding and Electronic Structure of Single Atom Dopants in Graphene with Electron Energy Loss Spectroscopy. *Nano Letters* **2013**, *13*, 4989-4995.

(24) Si-Vacancy Complexes in Graphene.(dataset). https://doi.org/10.25920/0xv3-8459.

(25) Vlassiouk, I.; Fulvio, P.; Meyer, H.; Lavrik, N.; Dai, S.; Datskos, P.; Smirnov, S. Large scale atmospheric pressure chemical vapor deposition of graphene. *Carbon* **2013**, *54*, 58-67.

(26) Garcia, A. G. F.; Neumann, M.; Amet, F.; Williams, J. R.; Watanabe, K.; Taniguchi, T.; Goldhaber-Gordon, D. Effective Cleaning of Hexagonal Boron Nitride for Graphene Devices. *Nano Lett.* **2012**, *12*, 4449-4454.

(27) Dyck, O.; Kim, S.; Kalinin, S. V.; Jesse, S. Mitigating e-beam-induced hydrocarbon deposition on graphene for atomic-scale scanning transmission electron microscopy studies. *J. Vac. Sci. Technol., B: Nanotechnol. Microelectron.: Mater., Process., Meas., Phenom.* **2017**, *36*, 011801.

(28) Kresse, G.; Joubert, D. From ultrasoft pseudopotentials to the projector augmented-wave method. *Physical Review B* **1999**, *59*, 1758-1775.

(29) Vanderbilt, D. Soft self-consistent pseudopotentials in a generalized eigenvalue formalism. *Physical Review B* **1990**, *41*, 7892-7895.

(30) Kresse, G.; Hafner, J. Norm-conserving and ultrasoft pseudopotentials for first-row and transition elements. *Journal of Physics: Condensed Matter* **1994**, *6*, 8245.

(31) Klimeš, J.; Bowler, D. R.; Michaelides, A. Van der Waals density functionals applied to solids. *Physical Review B* **2011**, *83*, 195131.

(32) Badrinarayanan, V.; Kendall, A.; Cipolla, R. SegNet: A Deep Convolutional Encoder-Decoder Architecture for Image Segmentation. *IEEE Transactions on Pattern Analysis and Machine Intelligence* **2017**, *39*, 2481-2495.

(33) Kingma, D. P.; Ba, J. Adam: A method for stochastic optimization. *arXiv preprint arXiv:1412.6980* **2014**.

(34) Lin, T.-Y.; Goyal, P.; Girshick, R.; He, K.; Dollár, P. Focal loss for dense object detection. *IEEE transactions on pattern analysis and machine intelligence* **2018**.

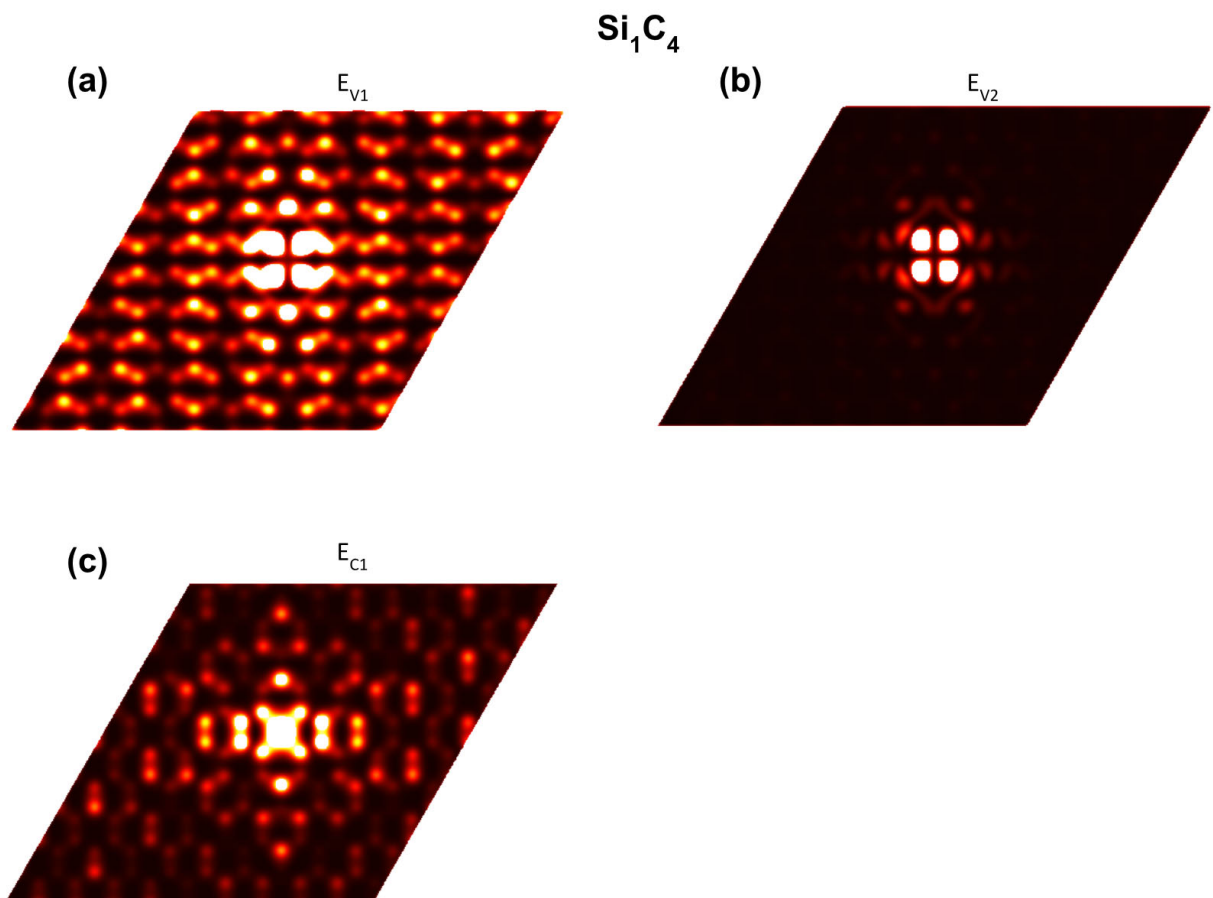

Figure S1. DFT-simulated distribution of electronic charge density for the bands below (a, b) and above (c) the Fermi level within the range from -0.5 eV to +0.5 eV of the single Si impurity with 4-fold coordination (see Figure 3b in the main text for details)

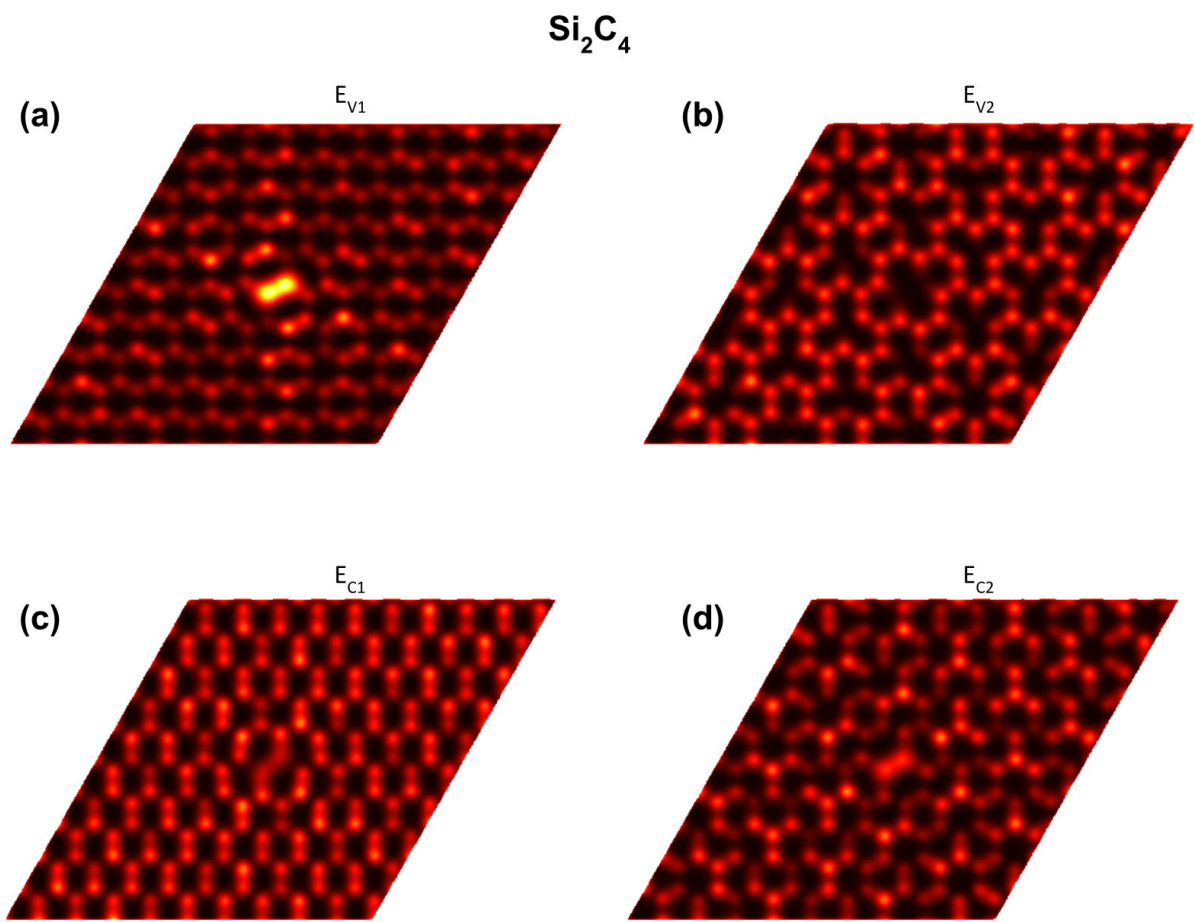

FIGURE S2. DFT-simulated distribution of electronic charge density for the bands below (a, b) and above (c, d) the Fermi level within the range from -0.5 ev to +0.5 eV of the Si dimer $Si_2C_4$ (see Figure 4a in the main text for details).

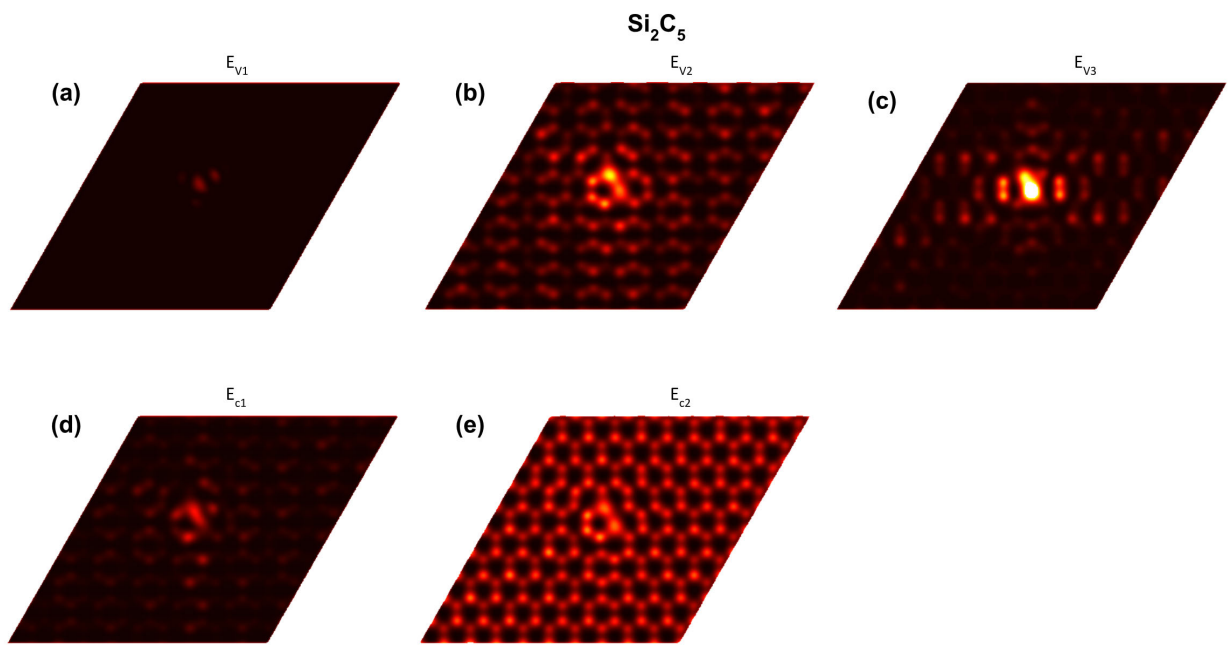

FIGURE S3. DFT-simulated distribution of electronic charge density for the bands below (a-c) and above (d, e) the Fermi level within the range from -0.5 ev to +0.5 eV of the Si dimer $Si_2C_5$ (see Figure 4b in the main text for details).

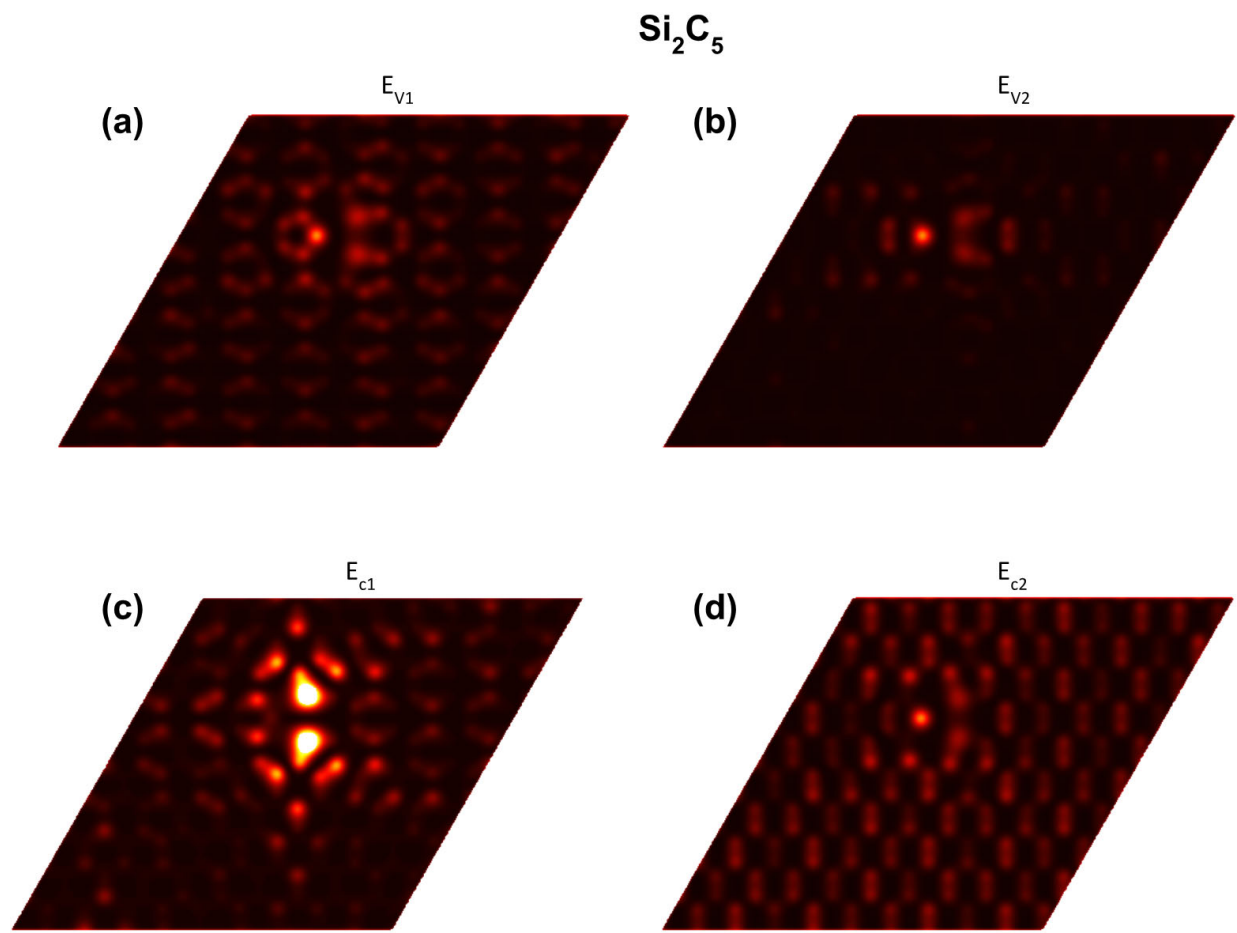

FIGURE S4. DFT-simulated distribution of electronic charge density for the bands below (a,b) and above (c, d) the Fermi level within the range from -0.5 ev to +0.5 eV of the Si dimer $Si_2C_5$ (see Figure 4c in the main text for details).

# Deep convolutional neural networks for atom finding: Testing against high levels of noise

*Author: Maxim Ziatdinov*

```
In [0]: import numpy as np
        import cv2
        from scipy import ndimage
        from skimage.feature import blob_log
        import scipy.spatial as spatial
        import itertools
        from keras.models import load_model
        import matplotlib.pyplot as plt
        from matplotlib import gridspec
        %matplotlib inline
```

Upload validation image and corresponding ground truth coordinates. The image was simulated using Multislice algorithm (Barthel, J. Dr. Probe: A software for high-resolution STEM image simulation. Ultramicroscopy 2018, 193, 1-11)

```
In [0]: lattice = np.load('graphene_val_img.npy')
        lattice_coord_gt = np.load('graphene_val_coord.npy')
```

```
In [0]: y_gt, x_gt = lattice_coord_gt.T
        plt.figure(figsize = (5,5))
        plt.scatter(x_gt, y_gt, s = 2)
        plt.imshow(lattice)
```

Out[0]: <matplotlib.image.AxesImage at 0x7f1402db2160>

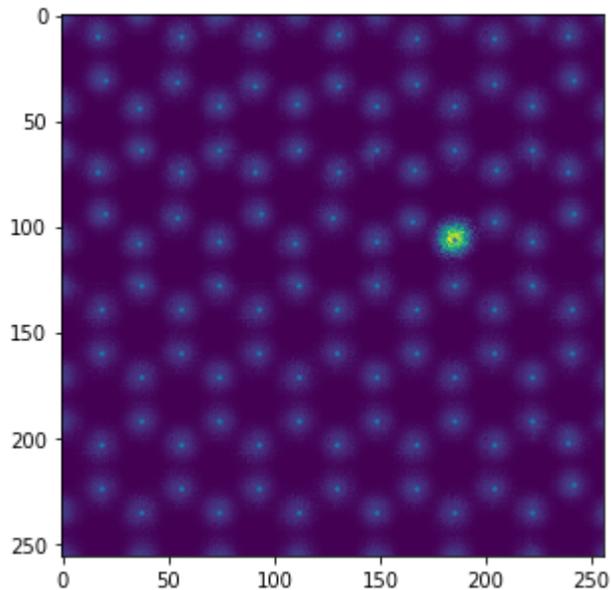

Corrupt image by different combinations of noise and blurring comparable to and higher than the experimental ones

```
In [0]: def make_blurred_image(image,s):
            image_n = ndimage.filters.gaussian_filter(image, s/20)
            return image_n

        def make_noisy_image(image,l):
            vals = len(np.unique(image))
            vals = (l/75) ** np.ceil(np.log2(vals))
            image_n_filt = np.random.poisson(image* vals) / float(vals)
            return image_n_filt
```

Do it for a single combination of noise and blurring values

```
In [0]: noise_level = 120
        blurring_level = 40

        lattice_bl = make_blurred_image(lattice, blurring_level)
        lattice_n = make_noisy_image(lattice_bl, noise_level)
```

Visualize results overlaid with ground truth coordinates

```
In [0]: plt.figure(figsize = (5,5))
        plt.scatter(x_gt, y_gt, s = 2, c = 'red')
        plt.imshow(lattice_n)
```

Out[0]: <matplotlib.image.AxesImage at 0x7f13fdca3128>

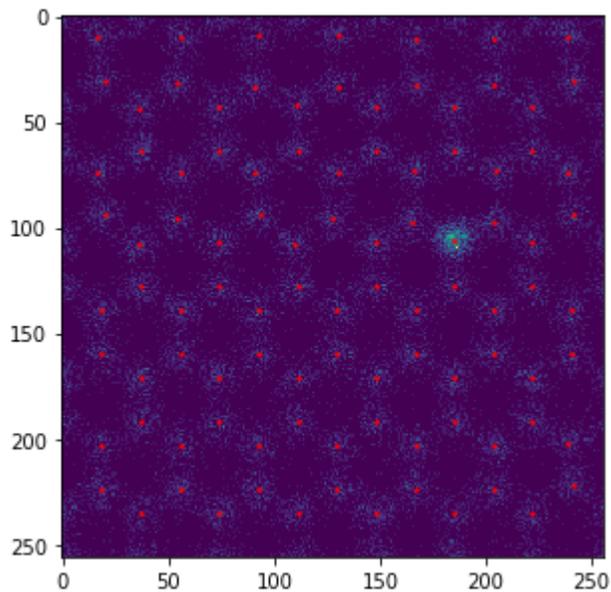

Now apply different combinations of noise and blurring to the data and store the resultant corrupted data as a dictionary.

```
In [0]: lattice_noise_test = {}
        for blurring_level in range(40, 150):
            for noise_level in range(105, 120):
                lattice_bl = make_blurred_image(lattice, blurring_level)
                lattice_n = make_noisy_image(lattice_bl, noise_level)
                lattice_noise_test[str(blurring_level)+'-'+str(noise_level)] = lat
        tice_n
```

Let's vizualize how the image looks like for medium, high and very high levels of noise.

```python
In [0]: noise = ['45-119', '90-114', '140-108']

        plt.figure(figsize=(20, 20))
        for i_, i in enumerate(noise):
            ax = plt.subplot(1, 3, i_+1)
            plt.imshow(lattice_noise_test[i], cmap = 'gray')
            ax.get_xaxis().set_visible(False)
            ax.get_yaxis().set_visible(False)
            ax.set_title("$\sigma'=$"+i.split('-')[0] + ", " + "$\lambda'=$" + i.split('-')[1])
            plt.tight_layout()
```

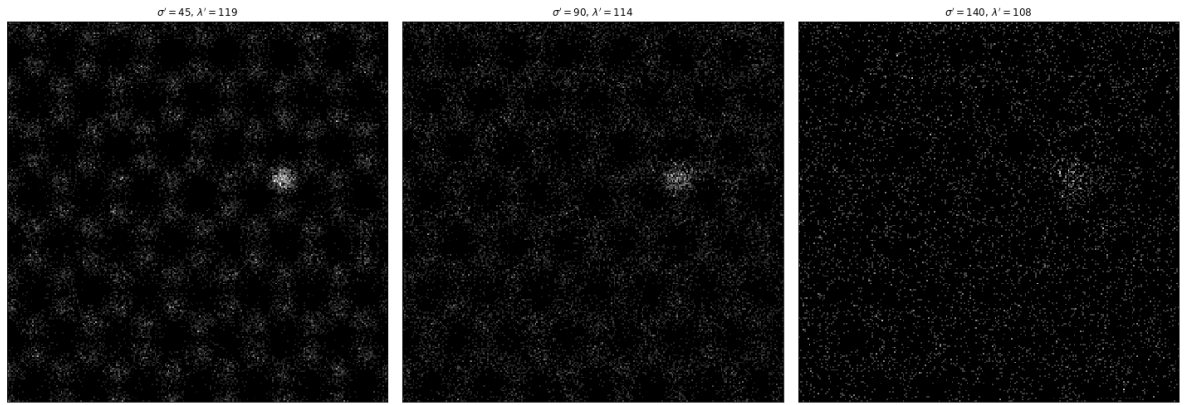

Save the produced data

```python
In [0]: np.save('lattice_noise_test-1.npy', lattice_noise_test)
```

Load a trained deep learning model

```python
In [0]: model = load_model('test-1.h5')
```

Decode the noisy data for every combination of noise and blurring.

```python
In [0]: def tf_format(image_data, image_size):
            '''Change image format to keras/tensorflow format'''
            
            image_data = image_data.reshape(image_data.shape[0], image_size[0], image_size[1], 1)
            image_data = image_data.astype('float32')
            image_data = (image_data - np.amin(image_data))/np.ptp(image_data)
            return image_data
```

```
In [0]: target_size = (256,256)
        lattice_noise_test_decoded = {}
        for k, v in lattice_noise_test.items():
            v = v.reshape(1, v.shape[0], v.shape[1])
            v = tf_format(v, target_size)
            decoded_img = model.predict(v)
            lattice_noise_test_decoded[k] = decoded_img[0,:,:,:]
```

Save the decoded results

```
In [0]: np.save('lattice_noise_test-1_decoded.npy', lattice_noise_test_decoded)
```

We will now analyze a deviation of the atomic coordinates in the decoded results from the ground truth coordinates (i.e. the coordinates that were used to generate a Multislice image). It is possible to run this section of a notebook without the previous one, so let's just load all the relevant data:

```
In [0]: lattice_noise_test = np.load('lattice_noise_test-1.npy')[()]
        lattice_noise_test_decoded = np.load('lattice_noise_test-1_decoded.npy')
        [()]
        lattice_coord_gt = np.load('graphene_val_coord.npy')
```

Let's see how well our model performs on an image where the level of noise is comparable to that seen in the experiment.

```
In [0]: k_i = '64-115'

        fig = plt.figure(figsize = (10, 10))
        gs = gridspec.GridSpec(1, 2, width_ratios=[1, 1])
        ax1 = plt.subplot(gs[0])
        ax1.imshow(lattice_noise_test[k_i], cmap = 'hot')
        ax1.axis('off')
        ax1.set_title("Validation data ($\sigma'=$"+k_i.split('-')[0] + ", "
                      + "$\lambda'=$" + k_i.split('-')[1] + ')')
        ax2 = plt.subplot(gs[1])
        ax2.imshow(lattice_noise_test_decoded[k_i], cmap = 'jet', alpha = 1)
        ax2.set_title('Network output (Softmax layer)')
        ax2.axis('off')
```

Out[0]: (-0.5, 255.5, 255.5, -0.5)

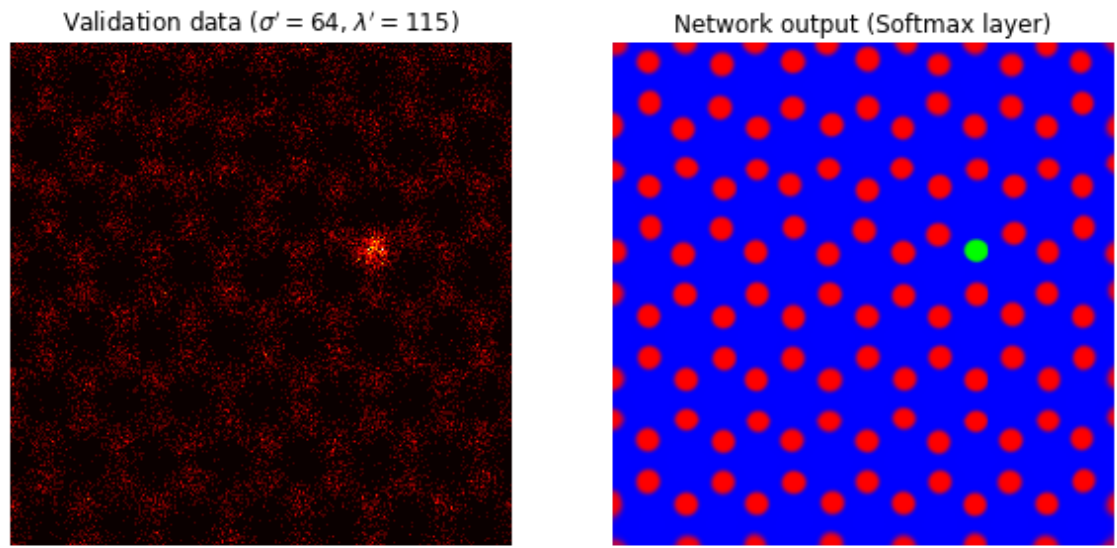

Now let's extract the deviation of atomic positions in the predicted results from ground truth atomic position for every considered combination of noise and blurring.

First define some helper functions:

In [0]:
```python
def coord_edges(coordinates, target_size, dist_edge):
    '''Remove image edges'''

    return [coordinates[0] > target_size[0] - dist_edge, coordinates[0] < dist_edge,
            coordinates[1] > target_size[0] - dist_edge, coordinates[1] < dist_edge]

def find_com(image_data):
    '''Find atoms via center of mass methods'''

    labels, nlabels = ndimage.label(image_data)
    coordinates = np.array(ndimage.center_of_mass(image_data, labels, np.arange(nlabels)+1))
    coordinates = coordinates.reshape(coordinates.shape[0], 2)
    return coordinates

def rem_coord(coordinates, target_size, dist_edge):
    '''Remove coordinates at the image edges'''

    coord_to_rem = [idx for idx, c in enumerate(coordinates) if any(coord_edges(c, target_size, dist_edge))]
    coord_to_rem = np.array(coord_to_rem, dtype = int)
    coordinates = np.delete(coordinates, coord_to_rem, axis = 0)
    return coordinates

def get_all_coordinates(decoded_imgs, target_size, method = 'LoG',
                       min_sigma = 1.5, max_sigma = 10, threshold = 0.8,
                       dist_edge = 3, verbose = 1):
    '''Extract all atomic coordinates in image via LoG or CoM methods & store data as a dictionary (key: frame number)'''

    d_coord = {}
    for i, decoded_img in enumerate(decoded_imgs):
        coordinates = np.empty((0,2))
        category = np.empty((0,1))
        for ch in range(decoded_img.shape[2]-1): # we assume that class 'background' is always the last one
            _, decoded_img_c = cv2.threshold(decoded_img[:,:,ch], threshold, 1, cv2.THRESH_BINARY)
            if method == 'LoG':
                coord = blob_log(decoded_img_c, min_sigma=min_sigma, max_sigma=max_sigma)
            elif method == 'CoM':
                coord = find_com(decoded_img_c)
            coord_ch = rem_coord(coord, target_size, dist_edge)
            category_ch = np.zeros((coord_ch.shape[0], 1))+ch
            coordinates = np.append(coordinates, coord_ch, axis = 0)
            category = np.append(category, category_ch, axis = 0)
        d_coord[i] = np.concatenate((coordinates, category), axis = 1)
    if verbose == 1:
        print("Atomic/defect coordinates extracted.\n")
    return d_coord

def find_nn(p1, coord, r = 20):
```

```python
        '''Find nearest neighbors for atom/impurity within specified radius
    r'''
    
        lattice_coord = np.copy(coord)
        lattice_coord_ = [p1]
        tree = spatial.cKDTree(lattice_coord)
        indices = tree.query_ball_point(lattice_coord_, r)
        indices = np.unique(list(itertools.chain.from_iterable(indices)))
        coord_new = np.empty((0,2))
        for i in indices:
            coord_new = np.append(coord_new, [lattice_coord[i]], axis = 0)
        return coord_new

def dist(p1, p2):
    '''Calculates distance between two points'''
    
    return np.sqrt((p1[0]-p2[0])**2 + (p1[1]-p2[1])**2)
```

Now find difference between predicted results and ground truth:

```python
In [0]: cf = (1/15.04) #pixels to angstroms conversion
        target_size = (256, 256)
        lattice_noise_test_coord = {}
        distance_d = {}
        distance_d_all = {}
        for k, v in lattice_noise_test_decoded.items():
            v = np.expand_dims(v, axis = 0)
            lattice_noise_coord = get_all_coordinates(v, target_size, method = 'Co
        M',
                                                      threshold = 0.5, dist_e
        dge = 5,
                                                      verbose = 0)
            coord = np.empty((0,3))
            d = np.empty((0,1))
            for c in lattice_coord_gt:
                c = c*cf
                c_ = lattice_noise_coord[0][:,0:2]*cf
                distance,index = spatial.KDTree(c_).query(c)
                d = np.append(d, distance)
                coord = np.append(coord, [lattice_noise_coord[0][index]], axis = 0
        )
            lattice_noise_test_coord[k] = coord
            distance_d_all[k] = d
            distance_d[k] = np.mean(d)
```

Now let's plot the results

```
In [0]: delta_r = []
        blur = []
        pois = []
        for k, v in distance_d.items():
            k = k.split('-')
            blur.append(int(k[0]))
            pois.append(int(k[1]))
            delta_r.append(v)
```

```
In [0]: X = np.array(blur)
        Y = np.array(pois)
        delta_r = np.array(delta_r)
```

```
In [0]: z_array = np.zeros((np.amax(Y)+1, np.amax(X)+1))
        for i,j,k in zip(X, Y, delta_r):
            z_array[j,i] = k*100 # from angstrom to picometers
```

```
In [0]: from mpl_toolkits.axes_grid1 import make_axes_locatable
        from mpl_toolkits.axes_grid1.inset_locator import zoomed_inset_axes
        from mpl_toolkits.axes_grid1.inset_locator import mark_inset
```

```
In [0]: fig, ax = plt.subplots(figsize=(12,36), dpi = 96)
        im = ax.imshow(z_array, cmap = 'jet')
        ax.scatter(float(k_i.split('-')[0]), float(k_i.split('-')[1]), marker =
        '+', c = 'red')
        ax.set_xlim(np.amin(X)-0.5,np.amax(X)+0.5)
        ax.set_ylim(np.amin(Y)-0.5,np.amax(Y)+0.5)
        ax.set_xlabel("$\sigma'$, a.u")
        ax.set_ylabel("$\lambda'$, a.u")
        ax.set_yticks([106, 111, 116])
        ax.set_title('Accuracy of finding atoms for different levels of noise and
         blurring')
        
        divider = make_axes_locatable(ax)
        cax = divider.new_vertical(size="20%", pad=0.8, pack_start=True)
        fig.add_axes(cax)
        clbar = fig.colorbar(im, cax=cax, orientation="horizontal")
        clbar.ax.set_xlabel('Position deviation, pm')
        plt.show()
```

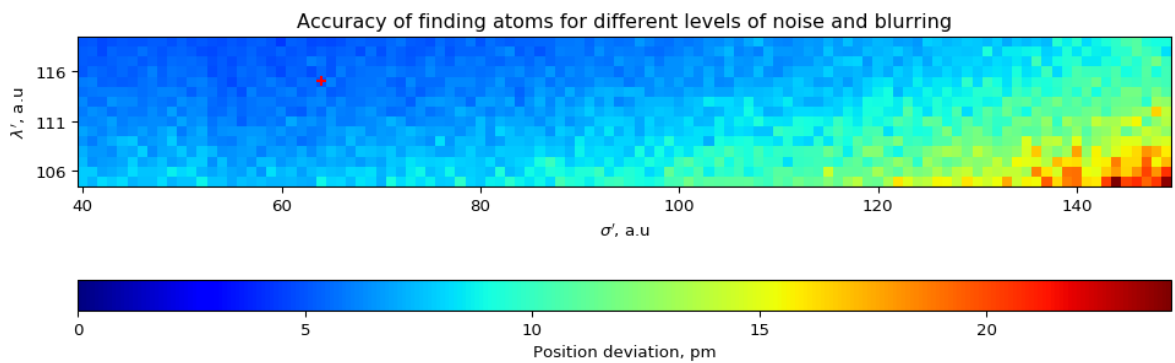

Below we show an example of the model output for data where the image resolution, number of atoms in image and the level of noise are comparable to those in current experiments. The deviation of atomic center positions from "true" positions for such data is mostly in 2-7 pm range, which is within the instrumental uncertainty of ~10 pm for detection of atomic position in STEM.

```python
In [0]: fig = plt.figure(figsize = (12, 12))
        gs = gridspec.GridSpec(1, 2, width_ratios=[1, 1])

        ax1 = plt.subplot(gs[0])
        y, x = lattice_noise_test_coord[k_i][:,0:2].T
        c = lattice_noise_test_coord[k_i][:,2].T
        ax1.imshow(lattice_noise_test[k_i], cmap = 'hot')
        im = ax1.scatter(x, y, s = 32, c = distance_d_all[k_i]*100,
                         cmap = 'Reds_r', vmin = 0)
        ax1.set_title("Decoded validation data ($\sigma'=$"+k_i.split('-')[0] + ", "
                      + "$\lambda'=$" + k_i.split('-')[1] + ')')
        ax1.axis('off')
        divider = make_axes_locatable(ax1)
        cax1 = divider.append_axes("bottom", size="5%", pad=0.3)
        clbar1 = fig.colorbar(im, cax=cax1, orientation = 'horizontal')
        clbar1.ax.set_xlabel('Position deviation, pm')

        hist, bin_edges = np.histogram(distance_d_all[k_i]*100, bins = 'auto')
        b_m = int(np.around(np.amax(bin_edges)))
        ax2 = plt.subplot(gs[1], aspect = np.ptp(bin_edges)/np.ptp(hist))
        ax2.hist(distance_d_all[k_i]*100, bins = range(1,b_m),
                 align = 'left', rwidth = 1, color = 'red')
        ax2.set_title('Deviation from ground truth coordinates')
        ax2.set_xlabel('Deviation, pm')
        ax2.set_ylabel('Count')
        plt.subplots_adjust(wspace = 0.35)

        plt.show(fig)
```

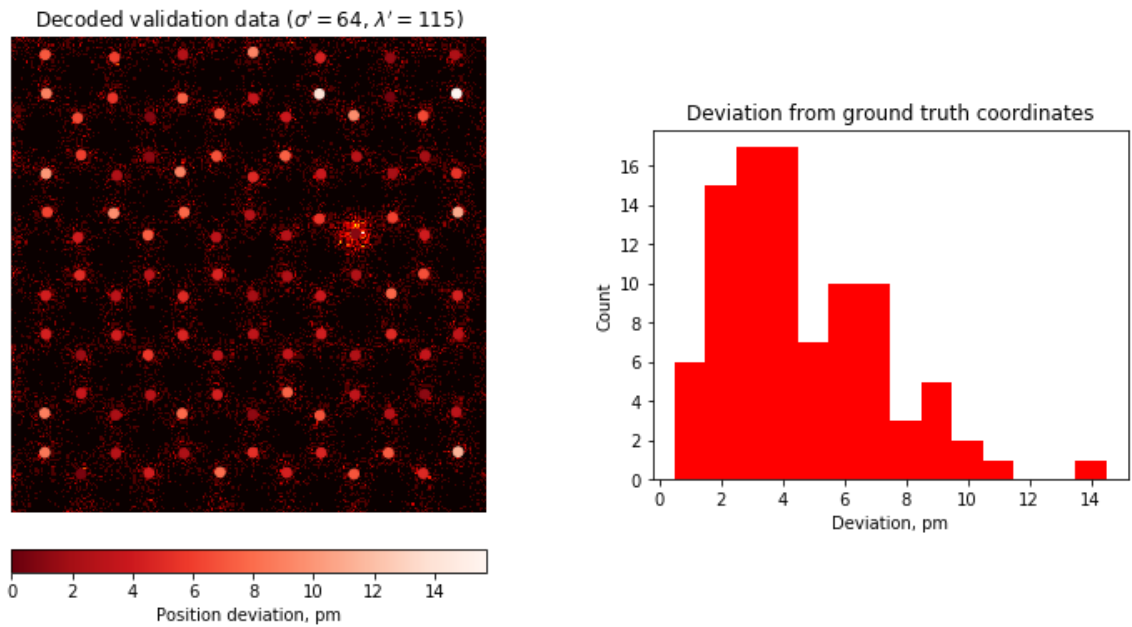

We can now do similar analysis for nearest neighbor distances. We first calculate all the nearest neighbor distances for ground truth coordinates:

```python
In [0]: def in_the_list(x, z):
            return any(np.array_equal(x, i) for i in z)
```

```python
In [0]: dist_all_gt = np.empty((0,1))
        nn_coord_all_gt = np.empty((0,2))
        coord = lattice_coord_gt
        coord = coord*cf
        for p1 in coord:
            nn = find_nn(p1, coord, 2)
            distance = np.array([])
            nn_coord = np.array([])
            for n in nn:
                d_i = dist(p1, n)
                distance = np.append(distance, d_i)
                nn_coord = np.append(nn_coord, n)
            
            distance_idx = np.concatenate((distance.reshape(-1, 1), nn_coord.reshape(-1, 2)), axis = 1)
            i = np.argsort(distance_idx[:,0])
            distance_idx_s = distance_idx[i,:]
            nn_coord = distance_idx_s[1:4,1:3]
            distance = distance_idx_s[1:4,0]
            
            
            for co, di in zip(nn_coord, distance):
                if di > 0 and in_the_list(co, nn_coord_all_gt) == False and in_the_list(p1, nn_coord_all_gt) == False:
                    nn_coord_all_gt = np.append(nn_coord_all_gt, [co], axis = 0)
                    dist_all_gt = np.append(dist_all_gt, di)
```

```python
In [0]: dist_gt_mean = np.mean(dist_all_gt)
```

Now calculate distances for decoded noisy data:

```python
In [0]: distance_d = {}
        distance_mean_d = {}
        for k, v in lattice_noise_test_decoded.items():
            v = np.expand_dims(v, axis = 0)
            coord = get_all_coordinates(v, target_size, method = 'CoM',
                                        threshold = 0.5, dist_edge = 5,
                                        verbose = 0)
            coord = coord[0][:,0:2]
            coord = coord*cf
            dist_all = np.empty((0,1))
            nn_coord_all = np.empty((0,2))
            for p1 in coord:
                nn = find_nn(p1, coord, 2)
                distance = np.array([])
                nn_coord = np.array([])
                for n in nn:
                    d_i = dist(p1, n)
                    distance = np.append(distance, d_i)
                    nn_coord = np.append(nn_coord, n)

                distance_idx = np.concatenate((distance.reshape(-1, 1), nn_coord.r
        eshape(-1, 2)), axis = 1)
                i = np.argsort(distance_idx[:,0])
                distance_idx_s = distance_idx[i,:]
                nn_coord = distance_idx_s[1:4,1:3]
                distance = distance_idx_s[1:4,0]

                for co, di in zip(nn_coord, distance):
                    if di > 0 and in_the_list(co, nn_coord_all) == False and in_th
        e_list(p1, nn_coord_all) == False:
                        nn_coord_all = np.append(nn_coord_all, [co], axis = 0)
                        dist_all = np.append(dist_all, di)
                distance_d[k] = dist_all
                distance_mean_d[k] = abs(np.mean(dist_all) - dist_gt_mean)
```

Plot the results:

```python
In [0]: delta_r = []
        blur = []
        pois = []
        for k, v in distance_mean_d.items():
            k = k.split('-')
            blur.append(int(k[0]))
            pois.append(int(k[1]))
            delta_r.append(v)
```

```python
In [0]: X = np.array(blur)
        Y = np.array(pois)
        delta_r = np.array(delta_r)
```

In [0]:
```python
z_array = np.zeros((np.amax(Y)+1, np.amax(X)+1))
for i,j,k in zip(X, Y, delta_r):
    z_array[j,i] = k*100 # transform to pm
```

In [0]:
```python
from mpl_toolkits.axes_grid1 import make_axes_locatable

fig, ax = plt.subplots(figsize=(12,36), dpi = 96)
im = ax.imshow(z_array, cmap = 'jet')
ax.scatter(float(k_i.split('-')[0]), float(k_i.split('-')[1]), marker = '+', c = 'red')
ax.set_xlim(np.amin(X)-0.5,np.amax(X)+0.5)
ax.set_ylim(np.amin(Y)-0.5,np.amax(Y)+0.5)
ax.set_xlabel("$\sigma'$, a.u")
ax.set_ylabel("$\lambda'$, a.u")
ax.set_yticks([106, 111, 116])

divider = make_axes_locatable(ax)
cax = divider.new_vertical(size="20%", pad=0.8, pack_start=True)
fig.add_axes(cax)
clbar = fig.colorbar(im, cax=cax, orientation="horizontal")
clbar.ax.set_xlabel('Bond length deviation, pm')
plt.show()
```

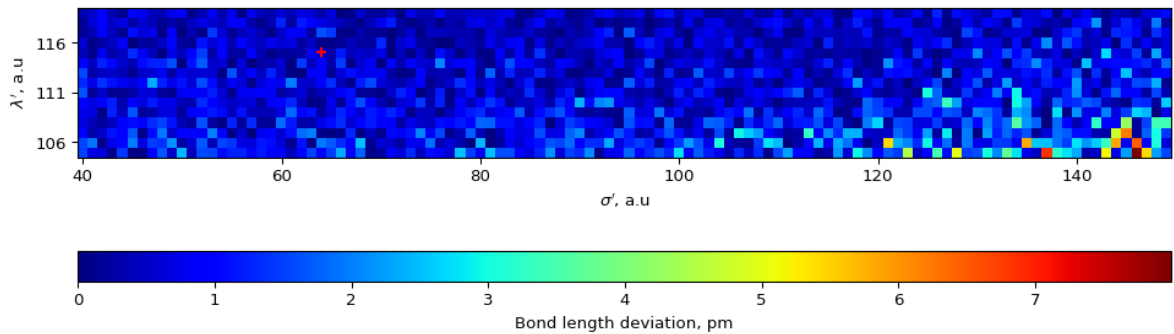

In [0]: